\documentclass[preprint,amsmath,amssymb,aps,prd,showpacs]{revtex4-1}

\usepackage{epsfig}
\usepackage{dcolumn}
\usepackage{bm}
\usepackage{tikz}
\usepackage{graphicx, graphics, color}
\usepackage{dcolumn}
\usepackage{bm}
\usepackage{hyperref}
\usepackage[mathlines]{lineno}

\newcommand{\be}{\begin{equation}}
\newcommand{\ee}{\end{equation}}
\newcommand{\ben}{\begin{eqnarray}}
\newcommand{\een}{\end{eqnarray}}

\newcommand{\de}{{\delta}}

\newcommand{\la}{{\lambda}}

\newcommand{\cO}{{\cal O}}

\newcommand{\cL}{{\cal L}}

\newcommand{\p}{\partial}
\newcommand{\na}{\nabla}

\newcommand{\Liek}{{\cal L}_{k^{\mu}}}
\newcommand{\trho}{{\tilde \rho}}

\newcommand{\tphi}{\tilde \phi}
\newcommand{\tpsi}{\tilde \psi}

\newcommand{\hSi}{{\hat \Sigma}}

\newcommand{\hg}{\hat g}
\newcommand{\hR}{\hat R}

\newcommand{\hna}{\hat \nabla}

\newcommand{\tG}{\tilde G}
\newcommand{\tH}{{\tilde H}}
\newcommand{\tg}{\tilde g}

\newcommand{\ep}{\epsilon}

\newcommand{\ga}{\gamma}

\newcommand{\tR}{{\tilde R}}

\newcommand{\tna}{\tilde \na}

\begin{document}

\title{Higher-dimensional electric-magnetic photon sphere uniqueness}
\author{Marek Rogatko} 
\email{rogat@kft.umcs.lublin.pl}
\affiliation{Institute of Physics, 
Maria Curie-Sklodowska University, 
20-031 Lublin, pl.~Marii Curie-Sklodowskiej 1, Poland}

\date{\today}

\begin{abstract}
In our paper we pay attention to the problem of uniqueness (classification) of higher-dimensional electro-magnetic
static, asymptotically flat, non-extremal solutions of multi-dimensional Einstein (n-2)-form gauge field gravity theory, which possess a photon sphere.
The photon sphere is uniquely described by its asymptotic data.
Conformal positive energy theorem is the key tool in the proof.
\end{abstract}

\maketitle

\section{Introduction}

 Recently the regions of spacetime in which photons form closed orbits, leading to the emergence of timelike hypersurfaces, the so-called photon spheres, on which
 the light angle is unrestrictedly high, attract attention to. The interests grow both from observational and theoretical points of view.
 Compact objects like black holes, neutron stars and wormholes being the short-cuts through spacetime, are encompassed by photon spheres.
 
 As far as the observational aspect is concerned, the monitoring  conducted by
 the Event Horizon Telescope collaboration (EHT), 
 of $M 87$ and Milky Way vicinities of supermassive black holes in their centres, are aimed at the verifications
of theoretical black hole characteristics \cite{vir00}-\cite{eht2}, as well as, the possible revealing first traces of the new physics beyond the Standard Model.

The special features of photon spheres in four-dimensional spacetimes, which resemble black hole event horizon ones, constitute the other powerful tool in
conducting classification of the aforementioned spacetimes, having in mind their asymptotical charges. Thus, the use of the photon sphere concept 
comprises the alternative way of achieving the black hole uniqueness theorems \cite{heu96}-\cite{rog24}.

The other mathematical and geometrical aspects of these objects were elaborated both for static and stationary axisymmetric spacetimes \cite{gib16}-\cite{kob22}.
It happens that they will also constitute the key role in studies Penrose inequalities \cite{shi17}-\cite{yan20}.

The generalization of the photon sphere idea to the case of massive charged particle,
describing timelike hypersurfaces to which any wordline
of particles initially touching to, remains in the hypersurface in question,  has been given in \cite{kob22a}-\cite{bog24a}. It has been also shown \cite{bog24b} that
a four-dimensional static asymptotically flat spacetime containing a massive particle sphere is isometric to Schwarzschild one, and this leads to
the existence of manifold foliations sliced by a set of massive particle spheres which are responsible for various values of energies.

On the other hand, the unification schemes such as M/string theory describe Universe as a brane or defect emerged in higher dimensional theory \cite{hor}, i.e., eleven-dimensional
spacetime with boundaries, where ten-dimensional Yang-Mills gauge theory reside on, trigger the interest in higher dimensional black holes. Especially the problem of compact objects
classification, like black holes, wormholes,
has been actively considered \cite{gib02}-\cite{wormholes18}. 
The idea of membrane Universe is still waiting for experimental verification, at TeV-energy scales, at which small black holes may be produced in LHC.

Consequently, the generalizations of the uniqueness theorem for $n$-dimensional gravity, by means of photon sphere attitude, were also under inspection.
Namely, in Ref. \cite{ced21} the higher-dimensional problem of photon sphere and uniqueness of higher-dimensional Schwarzschild spacetime was elaborated, while
the photon sphere uniqueness for electro-vacuum $n$-dimensional spacetime was conducted in \cite{jah19}. 
Moreover, the studies of the so-called
{\it trapped photons}, i.e., photons which never pass the event horizon or escape towards spatial infinity, in the spacetime of higher-dimensional Schwarzschild-Tangherlini black hole, 
were paid attention to in Ref.  \cite{bud20}. On the other hand, studies of equipotential photon surfaces in the context of $n$-dimensional electro-static spacetime uniqueness theorem,
were discussed in \cite{ced23}.

In our paper we shall treat the problem of classification of static asymptotically flat solutions to 
Einstein $(n-2)$-form gauge field gravity theory
containing
a photon sphere, in higher dimensional spacetime.
In the light of the recent measurements of black hole magnetic field conducted by EHT \cite{dal18}-\cite{aki218}, it will be interesting to have a glimpse at the
influence of magnetic field on photon sphere in higher-dimensional generalization of the problem in question.

Our paper is organized as follows. In Sec. II we briefly describe the main features of 
Einstein $(n-2)$-form gauge field gravity theory
in $n$-dimensional spacetime,
considering static spacetime with asymptotically timelike Killing vector field. In Sec. III we pay attention to a single component photon sphere 
and elaborate its mean curvature and scalar curvature in Einstein $(n-2)$-gauge form gravity. Sec. IV will be devoted to the studies of functional dependence of lapse function
on electric and magnetic potentials emerging in the theory under inspection. It has been revealed that $(n-3)$ electric forms and magnetic one-forms multiplied by the normal vector to the photon sphere are constant on it, justifying
the assertion that the consider photon sphere has constant curvature.
In Sec. V, using conformal positive energy theorem, one envisages that the photon sphere in question
is uniquely described by its asymptotic data, comprising Arnowitt-Deser-Misner (ADM) mass, electric and magnetic charges.
Sec. VI concludes our investigations.

\section{Higher dimensional 
Einstein $(n-2)$-form gauge field gravity system}

This section will be devoted to the formulation of $n$-dimensional 
Einstein $(n-2)$-form gauge field gravity theory with $(n-2)$-gauge
form $F_{\mu_{1} \dots \mu_{n-2}}$,
 which action is provided by
\be
S = \int d^n x \sqrt{-{g}} \Big[ {}^{(n)}R - F_{(n-2)}^2 \Big],
 \label{act}
\ee
where ${g}_{\mu \nu}$ is $n$-dimensional metric tensor,
$F_{(n-2)} = dA_{(n-3)} $ is $(n-2)$-gauge form field.
In the manifold under inspection we introduce asymptotically timelike Killing vector field.

Thus the corresponding line element of $n$-dimensional static spacetime, being subject to the
asymptotically timelike Killing vector field 
$k_{\alpha} = \Big( \frac{\p}{\p t }\Big)_{\alpha}$
and $N^{2} = - k_{\mu}k^{\mu}$, can be written in the following form:
\be
ds^2 = - N^2 dt^2 + g_{i j}dx^{i}dx^{j}.
\label{met}
\ee
On the other hand, the energy momentum tensor of the $(n-2)$-gauge form 
$T_{\mu \nu} = - \frac{\de S}{ \sqrt{- g}\de g^{\mu \nu}}$ yields
\be
T_{\mu \nu} = (n -2) F_{\mu i_{2} \dots i_{n-2}}
F_{\nu}{}{}^{i_{2} \dots i_{n-2}} - {g_{\mu \nu} \over 2}F_{(n-2)}^2.
\ee

In our consideration we shall take into account the asymptotically
flat spacetime, i.e., the spacetime will contain a data set 
$(\Sigma_{end}, ~g_{ij}, ~K_{ij})$ with gauge fields of $F_{(n-2)}$
such that $\Sigma_{end}$ is diffeomorphic to $\bf R^{n-1}$ minus 
a ball. The asymptotical conditions of the following forms should also be satisfied:
\ben \nonumber
\mid g_{ij} - \delta_{ij} \mid + r \mid \p_{a} g_{ij} \mid &+& \dots
+r^{m} \mid \p_{a_{1} \dots a_{m}} g_{ij} \mid + r \mid K_{ij}\mid + \dots
+r^{m} \mid \p_{a_{1} \dots a_{m}} K_{ij} \mid \le {\cal O}\Big( \frac{1}{ r} \Big), \\  \nonumber
\mid F_{i_{1} \dots i_{n-2}} \mid &+&
r \mid \p_{a} F_{i_{1} \dots i_{n-2}} \mid + \dots +
r^{m} \mid \p_{a_{1} \dots a_{m}}F_{i_{1} \dots i_{n-2}} \mid
\le {\cal O}\Big( \frac{1}{r^{2}} \Big).
\een
We define  {electric} $(n-3)$-form by the following expression:
\be
E_{i_{1} \dots i_{n-3}} = F_{i_{1} \dots i_{n-2}} k^{i_{n-2}},
\label{elec}
\ee
and {magnetic} $1$-form as
\be
B_{k} = \frac{1}{\sqrt{2 (n -2)!}}~ \ep_{k \mu i_{1} \dots i_{n-2}}
F^{i_{1} \dots i_{n-2}} k^{\mu}.
\label{mag}
\ee
Consequently, one can also introduce the rotation $(n-3)$-form of the stationary
Killing vector field $k_{\mu}$, given by
\be
\omega_{j_{2} \dots j_{n-2}} = \frac{(n-2) }{\sqrt{2 (n -2)!}}~
\ep_{j_{2} \dots j_{n-2} \mu \nu \gamma} k^{\mu} \na^{\nu} k^{\gamma}.
\label{tw}
\ee
Directly from equation (\ref{tw}) and definitions of {electric} and
{magnetic} forms, as well as, equations of motion for $(n-2)$-gauge form,
we find equations of motion for {magnetic} $1$-form $B_{k}$
\be
\na^{a}\Big( \frac{B_{a}}{ N^2 } \Big) = \frac{E^{i_{2} \dots i_{n-2}}~
\omega_{i_{2} \dots i_{n-2}} }{ N^4},
\label{bbb}
\ee
and similarly for {electric} $(n-3)$-form $E_{i_{1} \dots i_{n-3}}$
\be
\na^{j_{1}} \Big( \frac{E_{j_{1} j_{2}  \dots j_{n-3}} }{N^2}\Big) = - \frac{{B^{a}~\omega_{a j_{2} \dots j_{n-2}} }}{ N^4}.
\label{eee}
\ee
It turns out that an asymptotically flat, globally hyperbolic spacetime
is static, if
$
\omega_{j_{2} \dots j_{n-2}} = 0$, which imposes the condition that
$k^{\alpha} R_{\alpha [\beta }k_{\ga ]} = 0$.
The latter statement can be verified by the examination of the underlying equations of motion and the stationarity conditions for $F_{\mu_{1} \mu_{2} \dots \mu_{n-2}}$
with respect to the Killing vector field $k_{\mu}$, i.e.,
$
\Liek F_{\mu_{1} \dots \mu_{n-2}} = 0.
$
The direct calculations reveal that $R_{\alpha \beta}k^{\beta} = 0$ and it implies the statement in question.

Having in mind the exact form of the metric tensor (\ref{met}) 
on the hypersurface $\Sigma$ orthogonal to the Killing vector field $k_{\mu}$,
we assume that the only one non-trivial {electric}
component of the $(n-2)$-form $F_{\mu_{1} \dots
\mu_{n-2}}$ as $A_{0 1 \dots n-2} = \psi_F$ and for the
{magnetic} potential $B_{j} = \na_{j} \psi_B$.

On this account we get the following equations of motion for the underlying system:

\ben
\na_{i}\na^{i} N &=& \frac{(n - 3)^2 }{ N}\na_{i} \psi_F \na^{i} \psi_F
+ \frac{2 (n - 3)}{ (n - 2) N} \na_{i} \psi_B \na^{i} \psi_B, \\
\na_{i} \na^{i} \psi_F &=& \frac{1}{ N} \na \psi_F \na^{i} N, \\
\na_{i} \na^{i} \psi_B &=& \frac{1}{ N} \na_{i} \psi_B \na^{i} N, \\
{}^{(n-1)} R_{ij} &=& \frac{\na_{i} \na_{j} N}{N}
- \frac{n - 2}{ N^2} \na_{i}\psi_F \na_{j} \psi_F
+ \frac{g_{ij} }{N^2} \na_{i}\psi_F \na^{i} \psi_F \\ \nonumber
&-& \frac{2}{ N^2} \na_{i}\psi_B \na_{j} \psi_B + \frac{2 g_{ij}}{  (n - 2) N^2}
\na_{i} \psi_B \na^{i} \psi_B.
\een
The covariant derivative 
with respect to 
$g_{ij}$ is denoted by $\na$,
while ${}^{(n - 1)}R_{ij}(g)$ is the Ricci tensor defined on 
the hypersurface $\Sigma$.

\section{Photon sphere in higher dimensional theories}
 In our paper we shall deal with the problem of a single component 
 static photon surface ${}^{(n-1)} P$,
  being a timelike hypersurface embedded in 
 $n$-dimensional spacetime, for which any null geodesics which is initially tangent remains tangent during the time evolution of its existence.

In order to define a photon sphere one should take into account the notion of a lapse function $N$. 
 Consequently ${}^{(n-1)}P$ is the photon sphere if
 the lapse function $N$ is constant on ${}^{(n-1)}P$ and the additional requirement on fields existing in the theory under considerations, of being normal to it, is satisfied.
(see e.g., \cite{yaz15,yaz15b,yaz16}).

Moreover, as in four-dimensional case, one supposes that the lapse function in question regularly foliates the considered spacetime manifold outside the photon sphere, i.e., 
one has that  $1/\rho^2 = \na_a N \na^a N \ne 0$, outside the considered photon sphere.
$\na_b$ refers to the
 metric of $g_{ab}$ (\ref{met}).

 Consequently one defines the electric and magnetic static system, composed of\\
 $(M^{n-1}, ~{}^{(n-1)}g_{ij},~N, ~\psi_F,~\psi_B)$,
 as being a time slice of the following manifold
 $(R \times M^{n-1} \rightarrow - N^2 dt^2 + {}^{(n-1)} g_{ab} dx^a dx^b )$. By the photon sphere in question we shall
 consider a timelike hypersurface emebeded in $(R \times M^{n-1} \rightarrow - N^2 dt^2 + {}^{(n-1)} g_{ab} dx^a dx^b )$ under the auxiliary condition that its
 totally umbilic embedding (i.e., its second fundamental form comprises pure trace) and the gradient of the lapse function, as well as,  electric form is normal to photon sphere
 ${}^{(n-1)} P$ (see Sec. IV).

\subsection{Mean curvature}
In order to  find
 the main feature of the mean curvature of the photon sphere in the theory under inspection, we commence with the definition of 
the second fundamental form of photon sphere. It is given by the relation
$$ K_{ij} = \frac{1}{n-1} \Theta h_{ij},$$
where we denote the expansion of the unit normal to the photon sphere by $\Theta$ and $h_{ij}$ stands for the metric induced on it.

To proceed further, let us denote by $n_b$ the unit normal to ${}^{(n-1)} P$, while by $Y_b$ one sets the element of the tangent space $T{}^{(n-1)} P$. 
By means of the Codazzi equations, one obtains the following:
\be
\na_a K^{a}{}{}_b - \na_b K^m{}{}_m = R_{cd}~  n^d h^c{}{}_b,
\ee
and finally we arrive at the relation valid for all tangent vectors from the space $T{}^{(n-1)} P$.  It is provided by
\be
\frac{1}{n-1} \Theta_{, b} (2 - n)~Y^b = {}^{(n-1)} R_{b c}~ n^b Y^{c}.
\ee
The right-hand side of the above equation
can be rewritten 
using equations of motion and the definitions (\ref{elec}), (\ref{mag}). Namely in terms of $ F_{\mu_{1} \dots \mu_{n-2}} $ one gets
\be
R_{ab} = (n-2) F_{a \mu_{2} \dots \mu_{n-2}} F_{b }{}^{\mu_{2} \dots \mu_{n-2}} + \frac{g_{ab}}{n-2}  F_{\mu_{1} \dots \mu_{n-2}} F^{\mu_{1} \dots \mu_{n-2}}.
\ee
The main task is to find the form of the terms $F_{a \mu_{2} \dots \mu_{n-2}} F_{b }{}^{\mu_{2} \dots \mu_{n-2}} $ and
$F_{\mu_{1} \dots \mu_{n-2}} F^{\mu_{1} \dots \mu_{n-2}} $ with the use of the definitions for $B_k$ and $E_{i_1, \dots i_{n-3}}$.
After some algebra we obtain
\ben \label{fff1}
F_{\mu_{1} \dots \mu_{n-2}} F^{\mu_{1} \dots \mu_{n-2}} &=& (n-2) E_{\mu_1 \dots \mu_{n-3}}  E^{\mu_1 \dots \mu_{n-3}} - (n-2)! B_k B^k,
\\ \label{fff2}
F_{a \mu_{2} \dots \mu_{n-2}} F_{b}{}^{ \mu_{2} \dots \mu_{n-2}} &=&
(n-3) E_{a \mu_3 \dots \mu_{n-2}} E_{b}{}^{\mu_3 \dots \mu_{n-2}} - (n-3)! \Big( g_{ab} B^2 - B_a B_b \Big).
\een
We recall that by virtue of its definition $E_{\alpha_1 \dots \alpha_{n-2}}$ is normal to ${}^{(n-1)}P$.
 Having in mind the relations $k_{a} n^a = 0$ and $k_a Y^a = 0$, as well as, the fact that $\psi_B = \mu~\psi_F$ (see section IV), one gets 
that 
\be
 {}^{(n-1)} R_{b c}~ n^bY^{c} = 0.
\label{ric}
\ee
 It implies further that 
the following relation is provided:
\be
0 = \Big( 2 -n \Big) \frac{\Theta_{, b}}{(n-1)}.
\ee
This fact envisages that for arbitrary tangent vector $Y^a$, the mean curvature of the photon sphere under inspection is constant.

On the other hand, it can be seen that the fact that the following Lie derivative with respect to an arbitrary tangent vector to hypersurface ${}^{(n-2)}\Sigma$, 
given by
$\cL_{X_a} \Big( n^b~ {}^{(n-1)}\na_b N \Big)$ is equal to zero, autherizes the evidence that $n^b ~{}^{(n-1)}\na_b N$ is constant on the hypersurface in question.

\subsection{Scalar curvature}
Now we shall have a closer look at the scalar curvature of the photon sphere in static asymptotically flat spacetime,
 in the considered theory.
To commence with we recall the form of contracted Gauss equation in $n$-dimensional manifold, which yields
\be
{}^{(n-1)} R = {}^{(n)}R - 2 {}^{(n)} R_{ij}~ n^i n^j + \ep \Big( K_{a}{}^a K_{d}{}^{d} - K_{ab} K^{ab} \Big),
\ee
where in spacelike case $\ep =1$. On this account we arrive at the relation of the form
\be
{}^{(n-1)} R = \frac{T_{\mu}{}^\mu}{2 - n} - 2 T_{ab} ~n^a n^b + \frac{n-2}{n-1} \Theta^2.
\label{rrr}
\ee
The form of (\ref{rrr}), after using the exact relations for energy momentum components for tensor $F_{\mu_{1} \dots \mu_{n-2}}$, implies
\be
{}^{(n-1)} R =
\frac{1}{n-2} F_{\mu_{1} \dots \mu_{n-2}} F^{\mu_{1} \dots \mu_{n-2}} - 2 (n-2) F_{a \mu_{2} \dots \mu_{n-2}} F_{b}{}^{ \mu_{2} \dots \mu_{n-2}}.
\ee
In the next step, we implement the exact form of relations (\ref{fff1}) and (\ref{fff2}).
As can be seen,
in order to reveal that ${}^{(n-1)}R$ is constant, i.e., photon sphere under consideration has a constant scalar curvature, one ought to justify that
$E_{a \mu_3 \dots \mu_{n-2}} ~n^a $ and $B_a~ n^a$ (or alternatively $\na_a \psi_F~n^a,~\na_a \psi_B~n^a$ )
are constant on ${}^{(n-1)} P$.
Having in mind that $n^b ~{}^{(n-1)}\na_b N$ is 
 constant on  ${}^{(n-1)} P$, which was motivated at the end of the latter section, we have to prove that electric and magnetic
potentials are functions of $N$ (see next section for the proof). All these envisage that 
$E_{a \mu_3 \dots \mu_{n-2}} ~n^a $ and $B_a~ n^a$ are constant on the photon sphere in question and confirm
that $(n-1)$-dimensional photon sphere in 
Einstein $(n-2)$-form gauge field gravity theory
 has the constant curvature ${}^{(n-1)} R$.

\section{Functional dependence - lapse function electric and magnetic potentials}
Before we proceed to the main subject of the section, let us recall some useful in further considerations facts concerning the dependence
of magnetic and electric potentials.

To begin with, from the exact form of the twist vector $\omega_\alpha$  (\ref{tw}), one can conclude that in static spacetime with Killing vector field $k_\mu$,
it is equal to zero. 
Moreover if we take into account the relation valid for the Killing vector fields, i.e., $\na_{\alpha} \na_{\beta} k_{\ga}
= - R_{\alpha \beta \ga}{}{}{}^{\delta} ~k_{\delta}$, as well as, definitions given by the relations (\ref{elec}) and (\ref{mag}), it can be verified that the following equation can be achieved:
\be
\omega_{[j_{1} \dots j_{n-3}; b]} = - \alpha~E_{[j_{1} \dots j_{n-3}}~B_{b]},
\label{omeg}
\ee
where we set $\alpha = {\sqrt{2} (n - 2 )^{3/2} \over \sqrt{(n - 3)!}}$.

Due to the fact that we deal with the
static spacetime the right-hand sides of the relations (\ref{bbb}) and (\ref{eee}) 
are equal to zero. Moreover from the relation (\ref{omeg}) we obtain the following:
 \be
\ep^{\alpha_{1} \dots \alpha_{n-3}  \alpha_k}~\ep_{j_{1} \dots j_{n-3} k}~
E_{\alpha_{1} \dots \alpha_{n-3}}~B_{\alpha_k} = 0.
\ee

If we choose {electric} $(n-3)$-form as $E_{i_{1} \dots i_{n-3}} =
\delta^{0}_{i_{1}}~\delta^{0}_{i_{2}} \dots \delta^{m}_{i_{n-3}}~\na_{m} \psi_F$
and { magnetic} $1$-form $B_{k} = \na_{k} \psi_B$, it can be seen that $\psi_B = \mu ~\psi_F$.
Further, having in mind 
the asymptotic conditions for the potentials
$\psi_F \rightarrow 0$
and $\psi_B \rightarrow 0$, when $r \rightarrow \infty$
and the equation of motion for magnetic $B_{k}$ $1$-form in the static case, having the form as follows:
\be
\na^{a}\Big( {B_{a} \over N^2 }\Big) = 0,
\ee
one reveals that multiplying it by $\delta^{0}_{i_{1}}~\delta^{0}_{i_{2}} \dots \delta^{m}_{i_{n-3}}$
we obtain the relation
\be
\psi_B = \mu~\psi_F.
\label{magele}
\ee
where $\mu$ is constant.

Following the footsteps of derivations presented in Ref. \cite{isr68}, we define the coordinates on the submanifold
 $N=const,~t=const$ given by 
\be
g_{ab} dx^a dx^b = {}^{(n-2)}g_{ab} dy^a dy^b + \rho^2 dN^2,
\label{le}
\ee
where $\rho^{-1} = (\na_a N \na^a N)^{1/2}$ and $\na_a$ denotes the derivative with respect to $g_{ab}$ metric tensor.

Taking into account (\ref{magele}) and the definition of electric vector $E_{i_{1} \dots i_{n-3}}$ as purely spatial
$E_{i_{1} \dots i_{n-3}}k^{i_1} = 0$, 
one obtains the following relation:
\be
\frac{1}{\sqrt{{}^{(n-2)}g}} \frac{\p}{\p N} \Big[ \sqrt{{}^{(n-2)}g}~ \frac{\phi_F}{N} \Big] = - \frac{\na_a \Big( \rho~\na^a \psi_F \Big)}{N},
\label{elecvect}
\ee
with the definition
\be
\frac{\p \psi_F}{\p N} = \rho~\phi_F.
\ee

It happens that the gravitational equation which will be useful in the further derivations yield
\be
\frac{1}{\rho^2} \frac{\p \rho}{\p N} = K + \frac{(n-3) \rho}{N}~\Big[ \Big( n-3 \Big) + \frac{2 \mu^2}{n - 2} \Big]
\Big( \phi_F^2 + \na_a \psi_{F} \na^a \psi_F \Big),
\label{greq}
\ee
with $K = K_a{}^a$ being the extrinsic scalar curvature of the hypersurface $N=const$.

\subsection{Indentity}
Our main aim is to find the identity a'la Israel \cite{isr68}, comprising the arbitrary differentiable functions, say, $F$ and $G$, with the parameters of line element (\ref{le})
and relations for electric vector (\ref{elecvect}) and gravitational equation given by (\ref{greq}).

The functions in question, $F,~G$, will depend on $N,~\psi_F$, and one looks for the geneneral solution of the linear system of differentiable equations (which
constitutes the over-determined system), as combinations of particular solutions \cite{isr68}, as well as, we take into account
the integral conservation laws resulting from the considered identity. Consequently, taking all these into account, one obtains the identity as follows:
\ben \label{iquality}
\frac{1}{\sqrt{{}^{(n-2)}g}}
 \frac{\p}{\p N}  \Big[
 \sqrt{{}^{(n-2)}g} \Big( \frac{1}{N}
F(N, \tpsi ) \tphi &+& \frac{G(N, \tpsi )}{\rho} \Big) \Big] \\ \nonumber
= A~\rho~\Big( \tphi^2 + \na_a \tpsi \na^a \tpsi \Big) + B~\tpsi  &+& \frac{1}{\rho} \frac{\p G}{\p N}
- \frac{1}{N} \na_a \Big( F~\rho~\na^a \tpsi \Big).
\een
In the above relation (\ref{iquality}), we have set for $A$ and $B$ the following relations:
\ben
A &=& \frac{1}{N} \Big( G + \frac{\p F}{\p \tpsi} \Big) ,\\
B &=& \frac{1}{N} \frac{\p F}{\p N} + \frac{\p G}{\p \tpsi},
\een
while $\tpsi $ and $\tphi$ constitute the newly defined potentials of the forms
\ben
\tpsi &=& \sqrt{\Big( n-3 \Big) + \frac{2 \mu^2}{n - 2} }~ \psi_F,\\
\tphi &=& \sqrt{\Big( n-3 \Big) + \frac{2 \mu^2}{n - 2} }~ \phi_F.
\een

In order to achieve the integral conservation laws, we have to restrict our consideration to the case when
$ A = B = \frac{\p G}{\p N} = 0$. The general solutions of the above linear differential equations lead to the following particular solutions:
\be
F =1,~ G=0, \qquad F = (n-3) \tpsi,~ G=1, \qquad F = (n-3) \tpsi^2 - N^2,~G = (n-3) \tpsi.
\ee 
One can integrate the relation (\ref{iquality}), with respect to the all aforementioned values of functions $F$ and $G$, having in mind that the integral 
of two-dimensional divergence over a closed $N=const$ space disappears.
The two boundary surfaces $\Sigma_0$ and $\Sigma_\infty$ were taken into account, with the appropriate asymptotic conditions imposed on fields
and their characteristic features.

Having in mind the asymptotic behaviors of $\psi_F,~N$ given by
\be
N = 1 - \frac{M}{r^{n-2}} + \cO(1/r^{n-2}), \qquad \psi_F = \frac{Q_{(F)}}{(n-3) r^{n-3}} + \cO(1/r^{n-2}),
\ee
we arrive at the adequate conditions, when one approaches  $\Sigma_\infty$. In the case in question we have that
\ben
r^{n-3}  \psi_F & \rightarrow&  \frac{Q_{(F)}}{n-3},\\
r^{n-2} \phi_F &\rightarrow& -Q_{(F)},\\
\frac{\rho}{r^{n-2}} &\rightarrow& \frac{1}{(n-3)~M}.
\een
On the other hand, for $\Sigma_0$ hypersurface case, one has that $\phi_F = \cO(N),~\psi_{F;a} = \cO(N)$, while
on $\Sigma_0$ hypersurface both $\psi_F$ and $1/\rho$ are constant, as in four-dimensional case studied in Ref. 
\cite{isr67}.

Finally one arrives at the following:
\ben
\int_{\Sigma_0} dS ^{(n-2)}\Big( \frac{\phi_F}{N} \Big) &=& -\Omega_{(n-2)} (n-3) Q_{(F)},\\
(n-3) \Big[ \Big( n-3 \Big) &+& \frac{2 \mu^2}{n - 2} \Big] \psi_{(0) F}\int_{\Sigma_0} dS ^{(n-2)}
\Big( \frac{\phi_F}{N} \Big) + \frac{S_0}{\rho_0} = \Omega_{(n-2)} (n-3) M,\\ \nonumber
(n-3)\Big[ \Big( n-3 \Big) &+& \frac{2 \mu^2}{n - 2} \Big] \psi^2_{(0) F} \int_{\Sigma_0} dS^{(n-2)} \Big( \frac{\psi_F}{N} \Big) + (n-3) \frac{S_0}{\rho_0} \psi_{(0) F} = \\ 
&=&\Omega_{(n-2)} (n-3) Q_{(F)},
\een
where $S_0$ is the area  $\Sigma_0$ and $\Omega_{(n-2)}$ is the area of unit $(n-2)$-dimensional sphere.\\

All the above reveal the functional dependence among $N_0$ lapse function on $\Sigma_0$, ~$\psi_{(0) F}$ electric potential at
$\Sigma_0$ and the constant $\mu$ bounded magnetic and electric potentials. It yields
\be
\Big[ \Big( n-3 \Big) + \frac{2 \mu^2}{n - 2} \Big]\frac{ \psi_{(0) F}^2}{n-3} + (n-3) \psi_{(0) F} \frac{M}{Q_{(F)}} - 1 = N_0^2,
\label{funcn}
 \ee
 as was mentioned above $\psi_{(0) F}$ and $N_0$ are constant on the considered hypersurface and $\psi_F \rightarrow 0$, as $r \rightarrow \infty$. 

The equations (\ref{funcn}) is valid not only on the surface in question but also in all its exterior region. Namely, let us compose the divergence identity based on the 
above equation
\be
\frac{1}{2} \na^m \Bigg[\Big(-N^2 + \Big[ \Big( n-3 \Big) + \frac{2 \mu^2}{n - 2} \Big] \frac{\psi_F^2}{n-3} + \frac{(n-3) \psi_F M}{Q_{(F)}} -1\Bigg]~\theta_m  = N~\theta_m \theta^m,
\label{gauss}
\ee
where $\theta^m$ implies
\be
\theta^m = - \na^m N + \frac{1}{N} \Bigg[ \Big[ \Big( n-3 \Big) + \frac{2 \mu^2}{n - 2} \Big]
\psi_F~\frac{\na^m \psi_F}{n-3} + \frac{(n-3) M}{2 Q_{(F)}} \na^m \psi_F \Bigg].
\ee
In the next step, one applies the Gauss theorem to the relation (\ref{gauss}), taking into account the asymptotic behaviors of $N, ~\psi_F$, and the
fact that $N>0$ in the exterior region of {\it photon sphere}, one can draw a conclusion that $\theta^m = 0$. Fixing in this relation the integration constant as equal to 1,
we arrive at the equation expressing a functional dependence among electric/magnetic potentials and  $N$, which is provided by
\be
N^2 = \Big[ \Big( n-3 \Big) + \frac{2 \mu^2}{n - 2} \Big] \frac{\psi_F^2}{n-3} + \frac{(n-3)~M}{Q_(F)} \psi_F -1.
\ee
All the above prove the constancy of $E^{a i_{2} \dots i_{n-3}} n_a$ and $B_c n^c$ on the photon sphere ${}^{(n-1)}P$, revealing 
the constancy of its scalar curvature (see also the previous section).

\section{Uniqueness of n-dimensional photon sphere}
In this section we shall pay attention to the problem of the uniqueness of $n$-dimensional photon sphere with electric and magnetic charges. The main
tool we shall use to prove it, will be conformal positive energy theorem \cite{sim99}, widely implemented in classification of black holes, wormholes (see e.g. \cite{gib02}-\cite{wormholes18})
and also in four-dimensional photon sphere cases \cite{rog16,rog24}.  

As far as the conformal transformations are concerned, the first two are used in order to obtain regular hypersurfaces, on which total gravitational mass vanishes.
Then, the next ones were implemented in order to apply the conformal positive energy
theorem and to envisage that the static slice is conformally flat.

The main layout of the aforementioned theorem is two consider two asymptotically flat Riemannian $(n-1)$-dimensional manifolds equipped with
the metric tensors bounded with a conformal transformation
${}^{(\Psi)} g_{ab} = \Omega^2 ~{}^{(\Phi)} g_{ab}$, where $\Omega$ stands for the conformal factor. Moreover there is additional relation connected with the manifold masses
${}^{(\Psi)} m + \beta {}^{(\Phi)} m \ge 0$, under the auxiliary conditions imposed on their Ricci scalar curvature tensors ${}^{(\Psi)} R + \beta {}^{(\Phi)} R \ge 0$. 
The equality holds if and only the considered manifolds are flat.

Using the last conformal transformation reveals that the conformal flat spacetime can be rewritten in a form showing that Einstein $(n-2)$-gauge form equations of motion can be reduced to Laplace equation on $(n-1)$-dimensional Euclidean manifold,
which in turn enables us to conclude that the embedding of photon sphere is totally umbilical, hyperspherical (each component of photon sphere is a geometric sphere of a certain radius) and rigid, i.e., we can always locate one component of photon sphere at 
a certain point in the hypersurface, without loss of generality.

To commence with, 
let us define the quantities which are provided by the following relations, for electric potential:
\ben \label{ph1}
\Phi_{1} &=& \frac{1}{ 2} \Big[ N + \frac{1}{N} - \frac{(n - 2) {\psi_F}^2}{N}
\Big], \\
\Phi_{0} &=& \frac{\sqrt{n - 2}~ \psi_F}{ N},\\
\Phi_{-1} &=& \frac{1 }{ 2} \Big[ N - \frac{1 }{N} - \frac{(n - 2) {\psi_F}^2 }{ N}
\Big],
\een
and for the magnetic one
\ben
\Psi_{1} &=& \frac{1}{2} \Big[ N + \frac{1 }{N} - \frac{2 (n - 3) \psi_B^2}{N}
\Big], \\
\Psi_{0} &=& \frac{\sqrt{2 (n - 3)}~ \psi_B}{N},\\ \label{psi-1}
\Psi_{-1} &=& \frac{1 }{2} \Big[ N - \frac{1}{N} - \frac{2 (n - 3) \psi_B^2 }{N}
\Big].
\een
It can be revealed that the additional constraint relation can be achieved, i.e.,
\be
\Phi_{A} \Phi^{A} = \Psi_{A} \Psi^{A} = -1,
\label{rel}
\ee
where one defines the metric $\eta_{AB} = diag(1, -1, -1)$.

In the next step of our calculations we introduce the conformal transformation given by the following eqaution:
\be
\tg_{ij} = N^{\frac{2}{ n - 3}} g_{ij},
\ee
as well as, the symmetric tensors $\tG_{ij}$ and $\tH_{ij}$, bounded respectively with the potentials $\Phi_i$ and $\Psi_i$, defined by the equations (\ref{ph1})-(\ref{psi-1})
\ben \label{ggg}
\tG_{ij} &=& \tna_{i} \Phi_{-1} \tna_{j} \Phi_{-1} -
\tna_{i} \Phi_{0} \tna_{j} \Phi_{0} -
\tna_{i} \Phi_{1} \tna_{j} \Phi_{1},\\ \label{hhh}
\tH_{ij} &=& \tna_{i} \Psi_{-1} \tna_{j} \Psi_{-1} -
\tna_{i} \Psi_{0} \tna_{j} \Psi_{0} -
\tna_{i} \Psi_{1} \tna_{j} \Psi_{1},
\een
where we have denoted by $\tna_{i}$ the covariant derivative with respect to the metric tensor $\tg_{ij}$.

The above relations (\ref{ggg})-(\ref{hhh}) enable one to rewrite the equations of motion for the studied system in the form as follows:
\be
\tna^{2}\Phi_{A} = \tG_{i}{}{}^{i} \Phi_{A}, \qquad
\tna^{2} \Psi_{A} = \tH_{i}{}{}^{i} \Psi_{A},
\label{ppff}
\ee
where $A = - 1,~ 0,~ 1$. One recalls that the relations (\ref{ppff}) can be derived by varying the Lagrangian density \cite{mar02, hoe76, sim92}
\be
\cL = \sqrt{-\tg} \Big( \tG_{i}{}^{i} + \tH_{i}{}^{i} + \frac{ \tna^i  \Phi_{A} \tna_i  \Phi^{A}}{ \Phi_{A} \Phi^{A}} + \frac{ \tna^i  \Psi_{A} \tna_i  \Psi^{A}}{ \Psi_{A} \Psi^{A}} \Big),
\ee
with respect to $\tg_{ij},~ \Phi_{A}, ~ \Psi_{A}$, and taking into account the constraint relations (\ref{rel}).

On the other hand, it can be directly shown that
the Ricci tensor connected with the metric tensor $\tg_{ij}$ implies
\be
\tR_{ij} = \tG_{ij} + \frac{1}{n - 3}\tH_{ij}.
\label{rr}
\ee

To proceed further, one defines the next conformal transformations, which yield
\be
{}^{\Phi}g_{ij}^{\pm} = {}^{\phi}\omega_{\pm}^{\frac{2}{n - 3}} \tg_{ij},
\qquad
{}^{\Psi}g_{ij}^{\pm} = {}^{\psi}\omega_{\pm}^{\frac{2}{ n - 3}} \tg_{ij}.
\ee
Their conformal factors are given by
\be
{}^{\Phi}\omega_{\pm} = \frac{\Phi_{1} \pm 1 }{ 2}, \qquad
{}^{\Psi}\omega_{\pm} = \frac{\Psi_{1} \pm 1 }{2}.
\label{pf}
\ee
The above definitions enables us to conduct the construction presented , e.g., \cite{mas92}, where we build
manifolds $(\Sigma_{+}^{\Phi}, {}^{\Phi}g_{ij}^{+})$,
$(\Sigma_{-}^{\Phi}, {}^{\Phi}g_{ij}^{-})$, $(\Sigma_{+}^{\Psi}, {}^{\Psi}g_{ij}^{+})$, $(\Sigma_{-}^{\Psi}, {}^{\Psi}g_{ij}^{+})$,
which can be pasted
$(\Sigma_{\pm}^{\Phi}, {}^{\Phi}g_{ij}^{\pm})$ and 
$(\Sigma_{\pm}^{\Psi}, {}^{\Psi}g_{ij}^{\pm})$ across shared minimal boundaries. It leads to the construction of regular
regular hypersurfaces $\Sigma^{\Phi} = \Sigma_{+}^{\Phi} \cup \Sigma_{-}^{\Phi}$ and $\Sigma^{\Psi} = \Sigma_{+}^{\Psi} \cup \Sigma_{-}^{\Psi} $. 


Thus, the next point of studies will be connected with checking if that total
gravitational mass on hypersurfaces $\Sigma^{\Phi}$ and $\Sigma^{\Psi}$ 
is equal to zero. In order to perform this task, let us use the conformal positive mass theorem in higher dimensions \cite{gib02a,sim99} and consider
another conformal transformation given by
\be
\hg^{\pm}_{ij} = \bigg[ \bigg( {}^{\Phi}\omega_{\pm} \bigg)^2
 \bigg( {}^{\Psi}\omega_{\pm} \bigg)^{2 \la} \bigg]^{{1 \over (n-3)(1 + \la)}}\tg_{ij},
\label{pastctr}
\ee
where one sets $\la = 1 /n-3$.
Having in mind the relation (\ref{pastctr}), one finds that the Ricci curvature tensor connected with this conformally rescaled metric is provided by
\ben \label{ric}
(1 + \la) \hR &=& \Big[ {}^{\Phi}\omega_{\pm}^2~ {}^{\Psi}\omega_{\pm}^{2 \la} \Big]^{\frac{-1 }{ (n - 3)(1 + \la)}}
\Bigg( {}^{\Phi}\omega_{\pm}^{\frac{2}{ n - 3}}~ {}^{\Phi}R +
\la~ {}^{\Psi}\omega_{\pm}^{\frac{2}{ n - 3}}~{}^{\Psi}R \Bigg) \\ \nonumber
&+& \frac{\la }{1 + \la} \Big( \frac{n - 2}{ n - 3} \Big)
\Big( \hna _{i} \ln {}^{\Phi}\omega_{\pm} - \hna _{i} \ln {}^{\Psi}\omega_{\pm} \Big)  
\Big( \hna ^{i} \ln {}^{\Phi}\omega_{\pm} - \hna ^{i} \ln {}^{\Psi}\omega_{\pm} \Big).
\een

As can be checked by the direct calculations, the first term in brackets on the right-hand side of equation (\ref{ric}), is non-negative and yields
\ben
{}^{\Phi}\omega_{\pm}^{\frac{2}{ n - 3}}~{}^{\Phi}R +
\la~ {}^{\Psi}\omega_{\pm}^{\frac{2}{n - 3}}~{}^{\Psi}R &=& 
\Big( \frac{n - 2 }{ n - 3} \Big) \mid
\frac{\Phi_{0} \tna_{i} \Phi_{-1} - \Phi_{-1} \tna_{i} \Phi_{0} }{\Phi_{1} \pm 1 } \mid^2 \\ \nonumber
&+&
\frac{(n - 2)}{ (n - 3)^2} \mid \frac{ \Psi_{0} \tna_{i} \Psi_{-1} - \Psi_{-1} \tna_{i} \Psi_{0}}{\Psi_{1} \pm 1} \mid^2.
\een

Thus having in mind
 the conformal energy theorem, one concludes that
$(\Sigma^{\Phi}, {}^{\Phi}g_{ij})$, $(\Sigma^{\Psi}, {}^{\Psi}g_{ij})$ and
$(\hSi, \hg_{ij})$ are flat. In turn, this fact authorizes that the conformal factors satisfy
${}^{\Phi}\omega = {}^{\Psi}\omega$ and $\Phi_{1} = \Psi_{1}$, and moreover
$\Phi_{0} = const~ \Phi_{-1}$ and $\Psi_{0} = const~ \Psi_{-1}$. 

All the above reveal that the manifold $(\Sigma, g_{ij})$ is conformally flat, which enables us to rewrite $\hg_{ij}$ in a 
conformally flat form \cite{gib02,gib02a}, using the following definition
\be
\hg_{ij} = {\cal U}^{4 \over n-3} {}^{\Phi}g_{ij},
\label{gg}
\ee
with ${\cal U}$ denoted by  $({}^{\Phi}\omega_{\pm} N)^{-1/2}$.

Now we pay attention to the Ricci scalar $\hR$. It happens that its value is equal to zero implying that the
Einstein $(n-2)$-gauge form equations of motion can be reduced
to the Laplace equation on the $(n - 1)$ Euclidean manifold, i.e., 
$
\na_{i}\na^{i}{\cal U} = 0,
$
where $\na$ is the connection on a flat manifold. 
On this account one can define a local coordinate system with the line element given by
\be
{}^{\Phi}g_{ij} dx^{i}dx^{j} = \trho^{2} d{\cal U}^2 + {\tilde h}_{AB}dx^{A}dx^{B}.
\ee
The photon sphere in question will be located at some constant value of $\cal U$, with a radius described by a fixed value of $\rho$-coordinate \cite{gib02a}.

Having all the above in mind, we can define on hypersurface $\Sigma$ the metric line element can be written as 
\be
\hg_{ij}dx^{i}dx^{j} = \rho^2 dN^2 + h_{AB}dx^{A}dx^{B}.
\ee
In other words, the embedding of the photon sphere into Euclidean $(n-1)$-dimensional space will be totally umbilical. This fact implies \cite{kob69}
that such embedding is hyperspherical, i.e., each component of the photon sphere in question will constitute a geometric sphere of a certain radius.
Moreover, it turns out that the studied embedding is rigid \cite{kob69}, in the sense that one can always find one connected component of the photon sphere
at some fixed radius, without loss of generality.

On the other hand, if we have to do with one photon sphere at fixed radius, we have a boundary conditions of Dirichlet type for $\na_{i}\na^{i}{\cal U} = 0$, which indicates
that such solution must be spherically symmetric. 


 In the next step,
let us assume that ${\cal U}_{1}$ and ${\cal U}_{2}$ are  two solutions subject to the same of the boundary value problem and regularity.
The implementation of
the Green identity and integration over the volume element reveal the following:
\be
\bigg( \int_{r \rightarrow \infty} - \int_{\cal H} \bigg) 
\bigg( {\cal U}_{1} - {\cal U}_{2} \bigg) {\p \over \p r}
\bigg( {\cal U}_{1} - {\cal U}_{2} \bigg) dS = \int_{\Omega}
\mid \na \bigg( {\cal U}_{1} - {\cal U}_{2} \bigg) \mid^{2} d\Omega.
\ee
Due to the imposed boundary conditions, the surface integral vanishes implying that 
the volume integral is identically equal to zero, leading to the conclusion that the two discussed
solutions of Laplace equation subject to the Dirichlet boundary conditions are the same.
Summing it all up, the main conclusion of this section implies:\\
\noindent
{\bf Theorem}:\\
Let us suppose that $(M^{n-1}, ~{}^{(n-1)}g_{ij},~N, ~\psi_F,~\psi_B)$, is the asymptotic to the static, non-extremal,  
 $n$-dimensional Einstein $(n-2)$-gauge forms
$F_{\mu_{1} \dots \mu_{n-2}}$ black hole. The considered spacetime possesses photon sphere
$({}^{(n-1)}P, ~h_{ab}) \hookrightarrow (R \times M^{n-1} \rightarrow - N^2 dt^2 + {}^{(n-1)} g_{ab} dx^a dx^b )$, which can be considered as the inner boundary of
 $R \times M^{n-1} $. Denote respectively by $M,~Q_{(F)},~Q_{(B)}$, the ADM mass, total electric and magnetic charges, bounded with
$F_{\mu_{1} \dots \mu_{n-2}}$ gauge field.
Then, $(R \times M^{n-1} \rightarrow - N^2 dt^2 + {}^{(n-1)} g_{ab} dx^a dx^b )$ is isometric to the outer region of the photon sphere in the spacetime in question.


\section{Conclusions}
In our paper we have paid attention to the problem of uniqueness of photon sphere in $n$-dimensional black hole spacetime arising in Einstein $(n-2)$-gauge 
form $F_{\mu_{1} \dots \mu_{n-2}}$ gravity,
taking into account both electric and magnetic fields. One assumes the static case, i.e., we suppose the existence of asymptotically timelike
Killing vector field being orthogonal to the hypersurface of constant time. 

The functional dependence of the lapse function on electric and magnetic potentials was revealed, justifying the constancy of Ricci scalar curvature of the elaborated photon sphere.
Our main tool which has been implemented in the proof is conformal positive energy theorem which enables us to find that static asymptotically flat solutions of Einstein $(n-2)$- gauge form $F_{\mu_{1} \dots \mu_{n-2}}$ gravity, allowing the existence of photon sphere. The photon sphere in question is charcterized by ADM mass, electric and magnetic charges, under the assumption that the lapse function regularly foliate the considered spacetime.

\acknowledgments 
MR was partially supported by Grant No. 2022/45/B/ST2/00013 of the National Science Center, Poland.





\end{document}